\documentclass[prl,twocolumn,showpacs,preprintnumbers,amsmath,amssymb]{revtex4}
\usepackage{graphicx}
\usepackage{longtable}
\usepackage{dcolumn}
\usepackage{bm}
\usepackage{color}

\providecommand{\mr}{\mathrm}

\begin{document}

\title{Strength of the Effective Coulomb Interaction at Metal and Insulator Surfaces}

\author{Ersoy \c{S}a\c{s}{\i}o\u{g}lu}\email{e.sasioglu@fz-juelich.de}
\author{Christoph Friedrich}
\author{Stefan  Bl\"{u}gel}

\affiliation{Peter Gr\"{u}nberg Institut and Institute for
Advanced Simulation, Forschungszentrum J\"{u}lich and JARA,
52425 J\"{u}lich, Germany}

\date{\today}

\begin{abstract}

The effective on-site Coulomb interaction (Hubbard $U$) between
localized electrons at crystal surfaces is expected to be enhanced due 
to the reduced coordination number and reduced subsequent screening. By 
means of first principles calculations employing the constrained random-phase
approximation (cRPA) we show that this is indeed the case for
simple metals and insulators but not necessarily for transition
metals and insulators that exhibit pronounced surface states. In
the latter case, the screening contribution from surface states as
well as the influence of the band narrowing increases the electron
polarization to such an extent as to overcompensate the decrease
resulting from the reduced effective screening volume. The Hubbard
$U$ parameter is thus substantially reduced in some cases, e.g.,
by around 30\% for the ($100$) surface of bcc Cr.

\end{abstract}


\maketitle

The effective on-site Coulomb interaction (Hubbard $U$) between
localized electrons at surfaces of solids is expected to be
enhanced since the effective screening volume of the surface is
reduced with respect to the bulk. As a consequence, the electron
polarization decreases at the surface, which reduces the effect of
screening and gives rise to a larger $U$ value.  These arguments are 
underscored by interpolating between the Hubbard $U$ values of an isolated 
atom and an atom in a bulk solid, the former being 3 to 5 times larger than 
the latter \cite{Zaanen:90}.  Neither
experimental nor theoretical works have been reported so far that
would address the strength of the surface $U$ parameter
explicitly. However, a large number of phenomena observed in
solids indicates an enhancement of the $U$ at surfaces. For
instance, the metal-insulator transition at the surface of
correlated materials \cite{MIT}, the appearance of magnetism at
the surface of paramagnetic transition metals \cite{Goldoni}, and
the enhanced exchange splitting at the surface of 3\textit{d} ferromagnets
\cite{Exc} have been attributed to an increase of the correlation
strength, which is defined by the ratio $U/W$, where $W$ is
the bandwidth. In the theoretical description of surfaces, the $U$
is usually assumed to be unchanged \cite{Nolting,Liebsch,Eriksson}
so that the enhancement of correlation at the surface (S) with
respect to the bulk (B) is provided by the effective band
narrowing, i.e., $W_{\mr S} < W_{\mr B}$. In principle, depending
on the relative values of the surface $U$ and the bandwidth $W$,
the correlation strength can further increase or decrease even
below the bulk value. However, the latter case is considered to be
unlikely because it is believed that $U$ always increases at
surfaces. In this Letter, we show by means of first-principles
calculations that contrary to this conventional wisdom, this is not 
always the case. It decreases at  many transition-metal (TM) surfaces and  insulator 
surfaces with pronounced surface states, as a result of additional
screening channels that open up due to surface-related changes in the
electronic structure.

Recently, the calculation of the Hubbard $U$ parameter in solids
from first principles has been addressed by several authors
\cite{Kotani,Schnell,cLDA,cRPA,cRPA_Sasioglu,Werner,Wehling,Solovyev}. A number of
different approaches have been proposed and applied to the bulk
phase of various classes of materials. However, the effective
Coulomb interaction at a surface has been considered only within a
model Hamiltonian framework \cite{Ando}, so far. Reining and Del
Sole~\cite{Reining} performed model calculations to account for the 
contribution of the  surface states to the static electronic screening at the
Si($111$) surface. The authors showed that the
surface states give rise to a substantial enhancement of electron
screening at the surface, reducing the correlation strength.

The aim of this Letter is to determine the strength of the on-site
effective Coulomb interaction between localized electrons at metal
and insulator surfaces from first principles. To calculate the
Hubbard $U$ parameter we employ the constrained random-phase
approximation (cRPA) \cite{cRPA} within the full-potential linearized
augmented-plane-wave (FLAPW) method using maximally localized
Wannier functions (MLWFs) \cite{cRPA_Sasioglu,Max_Wan}. Our
calculations show that the Hubbard $U$ parameter is enhanced at simple 
metal (Na, Al) and  most insulator (SrTiO$_3$, NaCl) surfaces with
respect to bulk as expected. However, the situation is different
for TMs and insulators with pronounced surface states. 
For TM, both the interplay of the surface states 
and the effective band narrowing can give rise to a substantial reduction 
of the Hubbard $U$, while for insulators only the surface states 
are responsible for the reduction of $U$. For the ($100$) surface of 
bcc Cr  we obtain a 30\% reduction of the static $U$. Moreover, the 
frequency dependence $U(\omega)$ is markedly different from that of 
the bulk. For bcc Cr we find that, starting from $\omega=0$, the 
effective Coulomb interaction $U(\omega)$ increases monotonically with 
frequency, and at about 2~eV it exceeds the bulk value, which remains
basically constant in this interval.

We model the metal and insulator surfaces with slabs of 11 atomic
layers. Such slabs form a superlattice with 20~\AA{} of vacuum
separating them, with each slab possessing two ($100$) or ($110$)
symmetric surfaces. To discuss the chemical trends, additional modifications 
such as surface reconstruction and surface relaxation are at first not taken 
into account. We consider the 3\textit{d} TMs in their respective ground-state 
crystal structures except Sc, Ti, and Co (Mn), which are treated in the
fcc (bcc) structure. The ground-state calculations are 
carried out using the FLAPW method as implemented in the \texttt{FLEUR} 
code \cite{Fleur} with the GGA exchange-correlation potential as 
parameterized by Perdew \textit{et al.} \cite{GGA}. The MLWFs are 
constructed with the \texttt{Wannier90} code \cite{Wannier90,Fleur_Wannier90}. 
The effective Coulomb potential is calculated within the recently
developed cRPA method \cite{cRPA} implemented in the \texttt{SPEX} code
\cite{Spex} (for further technical details see
Refs.\,\onlinecite{cRPA_Sasioglu} and \onlinecite{Sasioglu}).

The cRPA approach offers an efficient way to calculate the
effective Coulomb interaction $U$ and allows to determine individual
Coulomb matrix elements, e.g., on-site, off-site, intra-orbital,
inter-orbital, and exchange as well as their frequency dependence.
The basic idea behind the cRPA \cite{cRPA} is to define an effective
interaction $U$ between the localized electrons by restricting the
screening processes to those that are not explicitly treated in
the effective model Hamiltonian. To this end, the full  RPA
polarization matrix $P$ is divided into  $P=P_{\mr l}+P_{\mr r}$,
where $P_{\mr l}$ includes only transitions between the localized
states, for which the Hubbard $U$ is to be calculated, and $P_{\mr r}$
is the remainder. Thus, the localized states are largely eliminated, 
and the screening is dominated by itinerant \textit{s} and \textit{p} 
states, which are well described by LDA and GGA. The cRPA is, therefore, 
-- even with these standarad xc potentials -- considered a reliable approach 
to calculate the Hubbard $U$ parameter and its frequency dependence 
\cite{Jepsen2010, Zhang2012}. Then, the frequency-dependent effective 
Coulomb interaction is given schematically by the matrix equation
$U(\omega) = [1-vP_{\mr r}(\omega)]^{-1}v$, where $v$ is the bare
Coulomb interaction. The static limit of the average diagonal
matrix element of $U(\omega \rightarrow 0)$ represented in a local
basis can then be regarded as the Hubbard $U$ parameter. The matrix
elements of the effective Coulomb potential $U$ in the
$\textrm{MLWF}$ basis are given by $U_{\mathbf{R}n_1 n_3;n_4
n_2}(\omega) = \iint w_{n_1\mathbf{R}}^{*}(\mathbf{r})
w_{n_3\mathbf{R}}(\mathbf{r})
U(\mathbf{r},\mathbf{r}^{\prime};\omega)w_{n_4\mathbf{R}}^{*}
(\mathbf{r}^{\prime})w_{n_2\mathbf{R}}(\mathbf{r}^{\prime})\:d^3r\:d^3r^{\prime}$,
where $w_{n\mathbf{R}}(\mathbf{r})$ is the MLWF at site
$\mathbf{R}$ with orbital index $n$ and
$U(\mathbf{r},\mathbf{r}^{\prime};\omega)$ is calculated
within the cRPA. We define the average on-site diagonal (direct
intra-orbital) and off-diagonal (exchange inter-orbital) matrix
elements of the effective Coulomb potential as
$U=L^{-1}\sum_{n}U_{\mathbf{R}nn;nn}$ and
$J=[L(L-1)]^{-1}\sum_{m,n (m\neq n)}U_{\mathbf{R}mn;nm}$, where
$L$ is the number of localized orbitals, i.e., one, three, and five for
\textit{s}, \textit{p}, and \textit{d} states, respectively.

\begin{figure}[t]
\begin{center}
\includegraphics[scale=0.55]{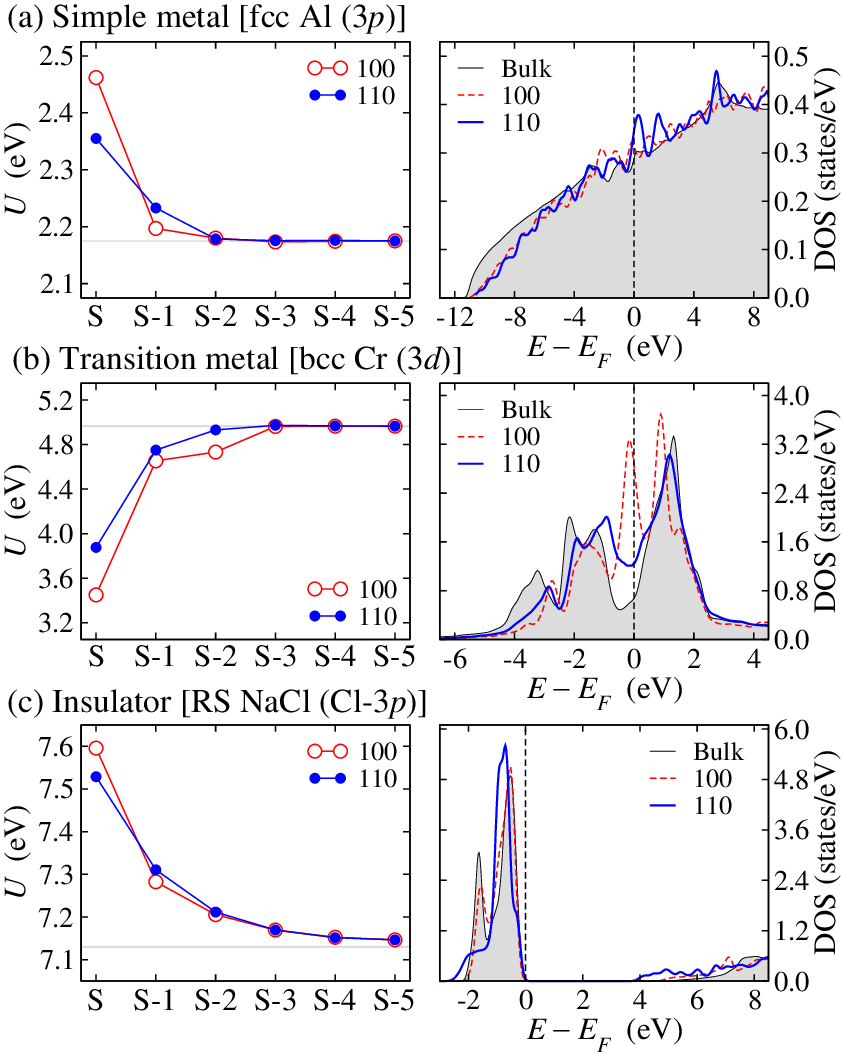}
\end{center}
\vspace*{-0.7cm} \caption{(Color online) (a) Left panel: layer
dependence of the Hubbard $U$ parameter for fcc Al. The $\mathrm{S}$ denotes
the surface layer in the slab model. Right panel: Total density of
states (DOS) for the (100) and ($110$) surface of fcc Al. For
comparison, the bulk DOS (shaded area) is included; (b) and (c) the
same for bcc Cr and rock-salt NaCl. } \label{layer_U}
\end{figure}

\begin{table}
\caption{Bulk and surface Hubbard $U$ and $J$ values for simple
metals and insulators. The corresponding orbitals for which the
$U$ and $J$ are calculated are given in parenthesis.}
\begin{ruledtabular}
\begin{tabular}{lcccccc}
&\multicolumn{2}{c}{Bulk} & \multicolumn{2}{c}{$100$ Surface} & \multicolumn{2}{c}{$110$ Surface} \\
                             & $U_{\mr B}$ & $J_{\mr B}$ &   $U_{\mr S}$ &  $J_{\mr S}$
			     &  $U_{\mr S}$ & $J_{\mr S}$  \\ \hline
Na  [3\textit{s}]                     & 1.39   &        &  1.50  &        &  1.47  &         \\
Al  [3\textit{p}]                     & 2.18   &  0.36  &  2.46  &  0.39  &  2.36  &  0.38   \\
MgO  [O-2\textit{p}]                  & 7.10   &  0.63  &  7.23  &  0.62  &  6.38  &  0.59   \\
NaCl [Cl-3\textit{p}]                 & 7.13   &  0.55  &  7.60  &  0.54  &  7.53  &  0.55   \\
SrTiO$_{\mathrm{3}}$ [Ti-3\textit{d}] & 3.34   &  0.37  &  3.62  &  0.39  &        &         \\
SrTiO$_{\mathrm{3}}$ [O-2\textit{p}]  & 4.42   &  0.56  &  4.79  &  0.55  &        &         \\
\end{tabular}
\label{table}
\end{ruledtabular}
\label{table}
\end{table}

In Fig.\,\ref{layer_U} we present the calculated layer dependence of
the Hubbard $U$ for three different systems: (a) for the $3p$ states 
of the simple metal fcc Al, (b) the $3d$ states of  the transition metal 
bcc Cr, and (c) the Cl-$3p$ states of the insulator NaCl in the rock-salt 
(RS) structure. As seen in cases (a) and (c), the Hubbard $U$
increases from the middle layer, where it is close to the bulk
value, to the surface layer as expected. However, we find an
unexpected behavior in the case of bcc Cr, where $U$ is
substantially reduced at the surface. This reduction is 30\% for
the open ($100$) and 20\% for the ($110$) surface. For
simplicity -- the magnetism of bcc Cr is quite complicated -- we
only consider the non-magnetic state here. Furthermore, the layer
dependence of $U$ is quite different in metals and insulators.
Because of the short screening length in metals, the Hubbard $U$
quickly assumes the bulk value in the former, as we go from the
surface toward the middle of the slab, while in the latter the
layer-by-layer convergence to the bulk value is much slower. 
On the other hand, the surface $J$ values, which are listed in
Table\,\ref{table} for various materials, only differ slightly from the
corresponding bulk values.

\begin{figure}[t]
\begin{center}
\includegraphics[scale=0.56]{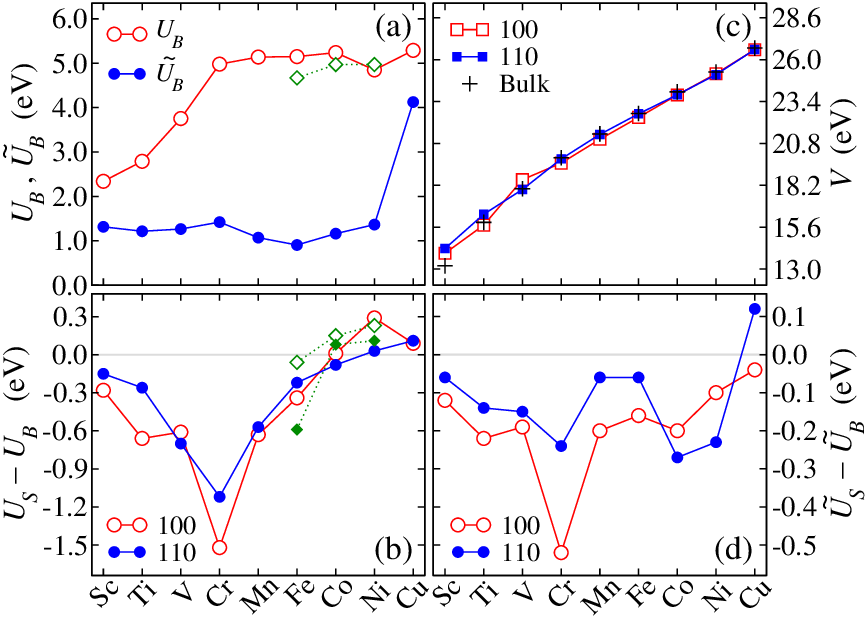}
\end{center}
\vspace*{-0.6cm} \caption{(Color online) (a) Partially screened
($U_{\mr B}$) and fully screened ($\tilde{U}_{\mr B}$) Coulomb interaction
for the bulk
3\textit{d} TM series in the non-magnetic state. With open
diamonds we show the $U$ for ferromagnetic Fe, Co, and Ni. (b)
The difference between the surface and bulk $U$ values. Open
[filled] diamonds show $U_{\mr S}-U_{\mr B}$ for the (100)  [(110)] surface
of ferromagnetic Fe, Co, and Ni. (c) Bare Coulomb interaction $V$
for bulk and surfaces. (d) The same as (b) for the fully screened
Coulomb interaction $\tilde U$.} \label{3d_Hubbard_U}
\end{figure}

We find that the surface electronic polarization and, as a consequence, the
Hubbard $U$ parameter is determined by two competing
effects: (i) the so-called dimensionality effect, which is due to
the reduced coordination number and hence the decrease of the
effective screening volume at the surface region. From the point
of view of classical electrostatics, this effect reduces the
electronic polarization at the surface leading to larger $U$ values. (ii)
Electronic structure effects, i.e., the appearance of surface states and 
the effective band narrowing.  This second effect gives rise to an enhancement
of the electronic polarization and, hence, to a decrease of $U$.
Depending on the strength of the two competing effects, the
effective Coulomb interaction at the surface can be enhanced as
well as reduced with respect to the bulk value. Qualitative
information on the influence  of the surface electronic structure
on the Hubbard $U$ parameter, leading to the second effect, can be
deduced from the polarization function, which is given by
\begin{eqnarray} \nonumber
P(\mathbf{r},\mathbf{r^{\prime}};\omega)&=2&\sum_{m}^{\mr {occ}}
\sum_{m^{\prime}}^{\mr {unocc}} \psi_{m}(\mathbf{r})
\psi_{m^{\prime}}^{*}(\mathbf{r})
\psi_{m}^{*}(\mathbf{r^{\prime}})
\psi_{m^{\prime}}(\mathbf{r^{\prime}}) \\ \nonumber &&\times \bigg
[\frac{1}{\omega - \Delta_{mm^{\prime}} +i\delta} -
\frac{1}{\omega + \Delta_{mm^{\prime}} -i\delta}\bigg ] \label{polar},
\end{eqnarray}
where $\psi_{m}(\mathbf{r})$ and $\epsilon_{m}$ are Kohn-Sham
eigenfunctions and eigenvalues, respectively, $\delta$ is a
positive infinitesimal, and 
$\Delta_{mm^{\prime}}=\epsilon_{m^{\prime}} - \epsilon_{m}$. For
zero frequency ($\omega=0$) the main contribution to the
polarization function comes from the states around the Fermi
energy. The smaller the energy difference between occupied and
unoccupied states the larger the contribution. The effective band 
narrowing in TMs tends to reduce
$\Delta_{mm^{\prime}}^{\textrm{S}}$ with respect to 
$\Delta_{mm^{\prime}}^{\textrm{B}}$, which has the effect of 
increasing the polarization. Additionally, the presence of 
surface states close to the Fermi level at the TM bcc ($100$) surfaces
make $\Delta_{mm^{\prime}}^{\textrm{S}}$ effectively smaller
resulting in a more efficient electronic polarization and, as a result, 
in substantially reduced $U$ and $\tilde{U}$ values as shown in 
Fig.\,\ref{3d_Hubbard_U} for the 3\textit{d} series, where $\tilde U$ 
stands for the fully screened Coulomb interaction within the RPA. 
As seen in Fig.\,\ref{layer_U}, the density of states 
(DOS) of the ($100$) surface of bcc Cr exhibits a pronounced peak 
that is caused by a surface state just below the Fermi level, which 
is also found in previous first-principles calculations \cite{SS_Cr}. 
This surface state contributes substantially to the polarization 
function and is mainly responsible for the 10\% stronger reduction of 
the $U$ value compared to the corresponding value for the ($110$) 
surface, where such a peak in the DOS is missing. On the other hand, 
in simple metals the surface electronic structure turns out to be
very similar to that of the bulk so that the dimensionality effect (i)
wins over the electron-structure effect (ii), giving rise to enhanced 
Hubbard $U$ parameters.

For most other 3\textit{d} metals, the Hubbard $U$ parameter is
reduced at the surface, too, but less than in the case of bcc Cr, and
for the late TMs the surface $U$ exceeds that of the bulk $U$. In
Fig.\,\ref{3d_Hubbard_U}(a) we present the partially screened $U_{\mr B}$
and fully screened $\tilde{U}_{\mr B}$ values for the bulk phases of 
the 3\textit{d} TMs in the non-magnetic state. Results for the 
ferromagnetic ground states of Fe, Co, and Ni are also included for
comparison. The difference between the surface and bulk $U$ and $\tilde U$
values, i.e., $U_{\mr S}-U_{\mr B}$ and $\tilde{U}_{\mr S}-\tilde{U}_{\mr B}$, 
is presented in Fig.\,\ref{3d_Hubbard_U}(b) and (d) for the ($100$) and ($110$)
surface. As seen, from Sc to Fe the Hubbard $U_{\mr S}$ is reduced at both
surfaces ($U_{\mr S}-U_{\mr B}<0$), and Co is at the border, in which the Hubbard $U$ assumes similar 
values in the bulk and at the surface. Only at the Ni and Cu surface the $U$ is
slightly larger. As for the difference of the fully screened Coulomb
interaction, i.e., $\tilde{U}_{\mr S}-\tilde{U}_{\mr B}$,  we obtain
a qualitatively similar behavior, but the relative reduction of
$\tilde{U}$ with respect to the bulk value is significantly larger.
This is attributed to the fact that, in contrast to the Hubbard $U$, 
screening effects stemming from $3\mathit{d} \rightarrow 3\mathit{d} $ transitions
contribute to the effective interaction, too. At the surface
these virtual transitions take place within the surface states, which
leads to very effective screening effects that give rise to the observed
reduction of the $\tilde U$. It is important to point out that the variations 
of $U$ seen in Fig.\,\ref{3d_Hubbard_U} are not caused by
different spreads of the Wannier function across the series, as can be 
seen from the bare Coulomb interaction $V$, which is presented in 
Fig.\,\ref{3d_Hubbard_U}(c). As seen, bulk and surface $V$ values monotonically 
increase from Sc to Cu, with no apparent difference between bulk and surface
values. We note that the Hubbard $U$ values depend only little on the bare 
Coulomb interaction $V$ because in metals  we are in the strong coupling limit, 
i.e., $v|P_{\mr l}| \gg 1$, and thus $U \simeq -P_{\mr l}^{-1}$. As for the 
surface $U_{\mr S}$ of the ferromagnets Fe, Co, and Ni,  the same discussion holds. 
So far, we have not taken into account the surface relaxation, which is
usually small ($<5\%$) for most of the 3d TMs \cite{SRelax1,SRelax2}. Only the 
bcc V ($100$) surface possesses a sizeable inward relaxation ($\sim$ 11\%), which 
results in a small change ($\sim$ 5\%) of the calculated surface $U$ and $\tilde{U}$ 
values.

In contrast to metals, the Hubbard $U$ at insulator surfaces is much 
more strongly affected by the presence of surface states than the band
narrowing. For example, the NaCl ($100$) surface does not exhibit
any surface states \cite{SS_NaCl}, and the $U$ is enhanced. The slight 
changes in the electronic structure of the ($110$) surface (the gap is 
smaller, see Fig.\,\ref{layer_U}) reduces the surface $U$ only very 
little so that it is still larger than the bulk value. The situation 
is similar for the SrTiO$_{\mathrm{3}}$ surface, for which the results 
are presented in Table\,\ref{table}. In MgO, on the other hand, both 
($100$) and ($110$) surfaces exhibit surface states, which effectively 
reduce the band gap. For the ($100$) surface this gap reduction is around 
0.6~eV \cite{SS_MgO}, while for the ($110$) surface we obtain a much 
stronger reduction, the surface states lie 3.3~eV below the conduction-band 
minimum, i.e., close to the middle of the  bulk GGA band gap, which 
amounts to 4.97~eV. The presence of the surface state strongly affects 
the screening properties with the consequence that the $U$ parameter is 
considerably reduced at the ($110$) surface, while it remains slightly 
above the bulk value in the case of the ($100$) surface. 
Analogously, localized states in the gap of an insulator stemming
from defects such as vacancies, interstitial or substitutional
impurities at surfaces lead also to a reduction of the surface
$U$, but it is expected to be local, i.e, only the Hubbard $U$
parameters of the defect atom and nearby atoms are affected.

\begin{figure}[t]
\begin{center}
\includegraphics[scale=0.58]{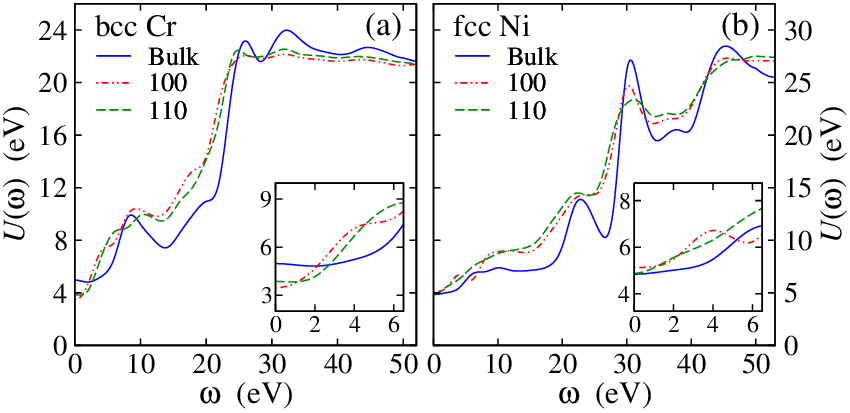}
\end{center}
\vspace*{-0.7cm} \caption{(Color online) (a) Frequency dependence
of the bulk and surface Hubbard $U$ parameter for bcc Cr. In the inset we
expand the low frequency region. (b) The same for fcc Ni.}
\label{Freq_dependence}
\end{figure}

Finally we discuss the frequency dependence of the surface $U(\omega)$ 
considering the non-magnetic state of bcc Cr and fcc Ni. The obtained 
results are presented  in Fig.\,\ref{Freq_dependence} and compared with the bulk
calculations. We see that, apart from the substantial reduction of the static
$U$ at the Cr surface, its frequency dependence is quite different from that of 
the bulk: $U_{\mr S}(\omega)$ increases monotonically with frequency, and at 
about 2~eV it crosses the $U_{\mr B}(\omega)$ curve towards larger values, while
$U_{\mr B}(\omega)$ stays almost constant between 0 and 4~eV.
For frequencies up to the plasmon frequency, $U_{\mr S}(\omega)$ 
is mostly larger than $U_{\mr B}(\omega)$. This behavior is not surprising because 
in the TMs  the 3\textit{d} states close to the Fermi level are much more affected
by the presence of the surface than the states at higher energies. Hence, 
with increasing frequency the polarization function becomes less sensitive to subtleties 
of the surface electronic structure, and at high frequencies the surface $U(\omega)$ 
tends to be larger than the bulk value due to the dimensionality effect.

In conclusion, by employing first-principles calculations in conjunction with 
the constrained random-phase approximation we have determined the strength of 
the Hubbard $U$ parameter at metal and insulator surfaces. We showed that  $U$ 
does not always increase at the surface as commonly expected. In fact, it decreases 
at most of the TM surfaces as well as insulator surfaces with pronounced surface states. 
We found that surface states and the effective band narrowing play an important role 
for the strength of the Hubbard $U$ at surfaces. The fact that the $U$ parameter can be 
made to increase as well as  decrease at surfaces offers new possibilities in designing 
materials with tunable correlations.

Fruitful discussions with  A. Liebsch, G. Bihlmayer, D. Wortmann,
A. Schindlmayr, and A. Lichtenstein are gratefully acknowledged.
This work has been supported by the DFG through the
Research Unit FOR-1346.


\begin{thebibliography}{99}

\bibitem{Zaanen:90}
J. Zaanen and G. A. Sawatzky, J. Solid State Chem. \textbf{88}, 8 (1990). 

\bibitem{MIT}
S.-K. Mo, J. D. Denlinger, H.-D. Kim, J.-H. Park, J. W. Allen, A.
Sekiyama, A. Yamasaki, K. Kadono, S. Suga, Y. Saitoh, T. Muro, P.
Metcalf, G. Keller, K. Held, V. Eyert, V. I. Anisimov, and D.
Vollhardt, Phys. Rev. Lett. \textbf{90}, 186403 (2003).

\bibitem{Goldoni}
A. Goldoni, A. Baraldi, G. Comelli, S. Lizzit, and G. Paolucci,
Phys. Rev. Lett. \textbf{82}, 3156 (1999).

\bibitem{Exc}
N. Kamakura, Y. Takata, T. Tokushima, Y. Harada, A. Chainani, K.
Kobayashi and S. Shin,  Europhys. Lett. \textbf{67}, 240 (2004).


\bibitem{Nolting}
M. Potthoff and W. Nolting, Phys. Rev. B \textbf{59}, 2549 (1999)


\bibitem{Liebsch}
A. Liebsch, Phys. Rev. Lett. \textbf{90}, 096401 (2003).


\bibitem{Eriksson}
A. Grechnev, I. Di Marco, M. I. Katsnelson, A. I. Lichtenstein,
J.Wills, and O.Eriksson, Phys. Rev. B \textbf{76}, 035107 (2007).



\bibitem{Kotani}
T. Kotani, J. Phys.: Condens. Matter \textbf{12}, 2413 (2000).


\bibitem{Schnell}
I. Schnell, G. Czycholl, and R. C. Albers, Phys. Rev. B \textbf{65},
075103 (2002); I. V. Solovyev and M. Imada, Phys. Rev. B \textbf{71}, 045103 (2005).



\bibitem{cLDA}
P.H. Dederichs, S. Bl\"ugel, R. Zeller, and H. Akai, Phys. Rev. Lett. \textbf{53}, 2512 (1984);
V. I. Anisimov and O. Gunnarsson, Phys. Rev. B \textbf{43}, 7570 (1991);
M. Cococcioni and S. de Gironcoli, Phys. Rev. B \textbf{71}, 035105 (2005);
K. Nakamura, R. Arita, Y. Yoshimoto, and S. Tsuneyuki, Phys. Rev. B \textbf{74},
235113 (2006).



\bibitem{cRPA}
F. Aryasetiawan, M. Imada, A. Georges, G. Kotliar, S. Biermann, and A. I.
Lichtenstein, Phys. Rev. B \textbf{70}, 195104 (2004); F. Aryasetiawan, K. Karlsson,
O. Jepsen, and U. Sch\"{o}nberger, Phys. Rev. B \textbf{74}, 125106 (2006);
T. Miyake, F. Aryasetiawan, and M. Imada Phys. Rev. B \textbf{80}, 155134 (2009).


\bibitem{cRPA_Sasioglu}
E. \c{S}a\c{s}{\i}o\u{g}lu, C. Friedrich, and S. Bl\"{u}gel, Phys.
Rev. B \textbf{83}, 121101(R) (2011).



\bibitem{Werner}
P. Werner, M. Casula, T. Miyake, F. Aryasetiawan, A. J. Millis, 
and S. Biermann, Nature Physics  \textbf{8}, 331  (2012).


\bibitem{Wehling}
T. O. Wehling, E. \c{S}a\c{s}{\i}o\u{g}lu, C. Friedrich, A. I. Lichtenstein, 
M. I. Katsnelson, and S. Bl\"{u}gel, Phys. Rev. Lett. \textbf{106}, 236805 (2011).


\bibitem{Solovyev}
I. V. Solovyev, Phys. Rev. Lett. \textbf{95}, 267205  (2005).



\bibitem{Ando}
T. Ando, A. B. Fowler, and F. Stern, Rev. Mod. Phys. \textbf{54},
437 (1982).

\bibitem{Reining}
L. Reining and R. Del Sole, Phys. Rev. B \textbf{38}, 12768
(1988).



\bibitem{Max_Wan}
N. Marzari and D. Vanderbilt, Phys. Rev. B \textbf{56}, 12847
(1997).



\bibitem{Fleur}
http://www.flapw.de


\bibitem{GGA}
J. P. Perdew, K. Burke, and M. Ernzerhof, Phys. Rev. Lett.
\textbf{77}, 3865 (1996).


\bibitem{Wannier90}
A. A. Mostofi, J. R. Yates, Y.-S. Lee, I. Souza, D. Vanderbilt,
and N. Marzari, Comput. Phys. Commun. \textbf{178}, 685 (2008).


\bibitem{Fleur_Wannier90}
F. Freimuth, Y. Mokrousov, D. Wortmann, S. Heinze, and S.
Bl\"{u}gel, Phys. Rev. B \textbf{78}, 035120 (2008).



\bibitem{Spex}
C. Friedrich, S. Bl\"{u}gel and A. Schindlmayr, Phys. Rev. B.
\textbf{81}, 125102 (2010).



\bibitem{Sasioglu}
E. \c{S}a\c{s}{\i}o\u{g}lu, A. Schindlmayr, C. Friedrich, F.
Freimuth and S. Bl\"{u}gel, Phys. Rev. B. \textbf{81}, 054434
(2010).


\bibitem{Jepsen2010}
K. Karlsson, F. Aryasetiawan, and O. Jepsen, Phys. Rev. B \textbf{81}, 245113 (2010) 


\bibitem{Zhang2012}
B-C. Shih, Y. Zhang, W. Zhang, and P. Zhang,  Phys. Rev. B \textbf{85}, 045132 (2012).




\bibitem{SS_Cr}
O. Yu. Kolesnychenko, G. M. Heijnen, A. K. Zhuravlev, R. de Kort,
M. I. Katsnelson, A. I. Lichtenstein, and H. van Kempen, Phys.
Rev. B \textbf{72}, 085456 (2005).



\bibitem{SRelax1}
J. L. F. DaSilva, C. Stampfl,  and M. Scheffler, Surf. Sci.
\textbf{600}, 703 (2006).


\bibitem{SRelax2}
G. Bihlmayer, T. Asada, and S. Bl\"{u}gel, Phys. Rev. B
\textbf{62}, R11937 (2000).


\bibitem{SS_NaCl}
B. Li, A. Michaelides, M. Scheffler, Phys. Rev. B \textbf{76},
075401  (2007).


\bibitem{SS_MgO}
U. Sch\"{o}nberger and F. Aryasetiawan, Phys. Rev. B \textbf{52},
8788 (1995).





\end{thebibliography}
\end{document}